**Analysis of characteristics of images acquired with a prototype clinical proton radiography system**

Christina Sarosiek[1], Ethan A. DeJongh[2], George Coutrakon[1], Don F. DeJongh[2], Kirk L. Duffin[3], Nicholas T. Karonis[3,4], Caesar E. Ordoñez[3], Mark Pankuch[5], Victor Rykalin[2], John R. Winans[3], James S. Welsh[6,7]

[1]Department of Physics, Northern Illinois University, DeKalb, IL 60115, USA
[2]ProtonVDA LLC, Naperville, IL 60563, USA
[3]Department of Computer Science, Northern Illinois University, DeKalb, IL 60115, USA
[4]Argonne National Laboratory, Data Science and Learning Division, Argonne, IL 60439, USA
[5]Northwestern Medicine Chicago Proton Center, Warrenville, IL 60555, USA
[6]Edward Hines Jr VA Medical Center, Radiation Oncology Service, Hines, IL 60141, USA
[7]Department of Radiation Oncology, Loyola University Stritch School of Medicine, Maywood, IL 60153, USA

## ABSTRACT

Purpose: Verification of patient specific proton stopping powers obtained in the patient's treatment position can be used to reduce the distal and proximal margins needed in particle beam planning. Proton radiography can be used as a pre-treatment instrument to verify integrated stopping power consistency with the treatment planning CT. Although a proton radiograph is a pixel by pixel representation of integrated stopping powers, the image may also be of high enough quality and contrast to be used for patient alignment. This investigation quantifies the accuracy and image quality of a prototype proton radiography system on a clinical proton delivery system.

Methods: We have developed a clinical prototype proton radiography system designed for integration into efficient clinical workflows. We tested the images obtained by this system for water-equivalent thickness (WET) accuracy, image noise, and spatial resolution. We evaluated the WET accuracy by comparing the average WET and rms error in several regions of interest (ROI) on a proton radiograph of a custom peg phantom. We measured the spatial resolution on a CATPHAN Line Pair phantom and a custom edge phantom by measuring the 10% value of the modulation transfer function (MTF). In addition, we tested the ability to detect proton range errors due to anatomical changes in a patient with a customized CIRS pediatric head phantom and inserts of varying WET placed in the posterior fossae of the brain. We took proton radiographs of the phantom with each insert in place and created difference maps between the resulting images. Integrated proton range was measured from an ROI in the difference maps.

Results: We measured the WET accuracy of the proton radiographic images to be ±0.2 mm (0.33%) from known values. The spatial resolution of the images was 0.6 lp/mm on the line pair phantom and 1.13 lp/mm on the edge phantom. We were able to detect anatomical changes producing changes in WET as low as 0.6 mm.

Conclusion: The proton radiography system produces images with image quality sufficient for pre-treatment range consistency verification.

## 1. INTRODUCTION

Proton radiation therapy delivers a conformal dose to a target volume in a patient. This requires precise knowledge and consistency of the patient's anatomy to ensure accurate dose delivery. Currently, medical physicists create proton treatment plans using x-ray CT images, and these plans assume that the patient anatomy remains constant during all treatment fractions. However, errors in the Hounsfield units (HU) to relative stopping power (RSP) conversion, unexpected changes in anatomy, or misalignments of the patient with respect to the proton beam can cause overdosing of healthy tissues surrounding the tumor and/or underdosing of the tumor volume[1]. Such dose differences can lead to serious short- or long-term side effects affecting the patient's quality of life post-treatment or treatment failure due to target underdosing.

Proton radiography measures the proton path integral of RSP, called water-equivalent path length (WEPL), through the patient in the proton beam's eye view. Reconstruction software bins each proton's WEPL into a 2D image of water-equivalent thickness (WET)[a] projected to a chosen plane, such as the isocenter, front, or rear tracker plane. A proton radiograph can therefore be useful in a clinical setting as a pre-treatment quality assurance device. In particular, proton radiography can be used to detect differences in integrated proton range that could arise from inter-fractional changes in anatomy before the patient is irradiated[2,3]. A digitally reconstructed proton radiograph from the treatment planning CT compared to a proton radiograph taken on the day of treatment would allow for detection of possible proton range errors prior to treatment. Additionally, proton radiographic images may be of high enough quality to be used for patient alignment[4,5]. However, in unusual situations, the HU to RSP conversion uncertainties, anatomical changes, and patient alignment errors may cancel in the proton radiograph and result in non-detection of the range error.

A simple range probe has a similar capability to detect range errors by shooting a single pencil beam through the tumor site and measuring the location of the Bragg peak in a multi-layered chamber behind the patient[6]. This method uses clinical quality assurance equipment and allows for fast implementation into clinical workflow. However, the range probe gives no spatial information and could not be used for patient alignment. Proton radiography requires more complex instrumentation and computer software but gives a full 2D image of the patient. There are two main approaches to proton radiography: particle-tracking and integrated. The integrated approach measures pencil beams of known locations with a suitable detector, such as a flat panel array, downstream of the object[7]. Knowledge of the incident energy and measurement of the residual energy of the protons allows for determination of the average WEPL of many protons in a pixel. This approach offers a reduction of detector complexity but results in poor spatial resolution. Particle-tracking proton radiography tracks the location and sometimes angle of individual protons upstream and downstream of the object[8]. This method requires more complex detector systems but allows for determination of individual proton trajectories using a most likely path (MLP) algorithm resulting in spatially resolved images.

ProtonVDA LLC developed a clinical prototype proton radiography system, shown in Figure 1, with design considerations to easily transition into a proton treatment room as described by DeJongh, et al[9]. The associated image reconstruction software, as described by Ordoñez, et al[10], produces clinical proton radiographic images with less than one minute of reconstruction time. In order for the clinical implementation to be successful, the proton radiographic images need to possess good and useful image quality while maintaining a low dose to the patient. A demonstration that proton radiography systems are ready for transition into the clinic requires characterization of proton radiography images of phantoms in terms of range as well as spatial resolution.

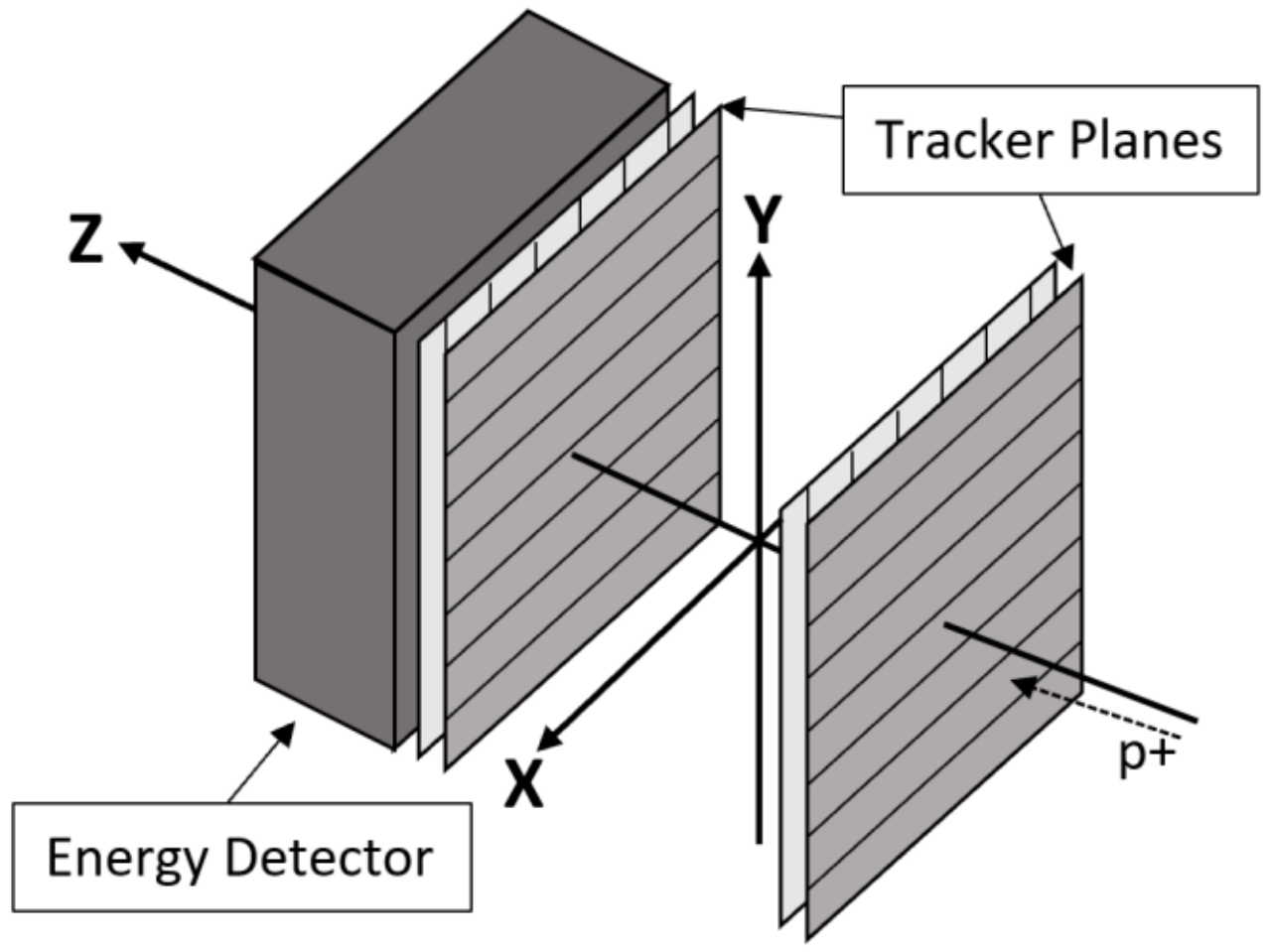

*Figure 1: Schematic of proton radiography system and coordinate system.*

In this paper, we report on the image quality of proton radiographic images reconstructed with data acquired with the ProtonVDA detector in a horizontal beam room at Northwestern Medicine Chicago Proton Center (NMCPC) and reconstructed with software implemented for prompt, online automatic reconstruction on a CPU/GPU computer and optimized with parallel processing. We imaged three phantoms to measure the WET accuracy and spatial resolution and we tested the ability to detect proton range differences that result from anatomical changes by interchanging inserts of known WET in a customized pediatric head phantom.

## 2. METHODS AND MATERIALS

### 2.1 Detector Hardware

The detector is comprised of one tracker plane upstream of the phantom, one tracker plane downstream of the phantom, and a compact residual energy detector. Each tracker plane has two layers of scintillating fibers to measure the X and Y positions of individual protons upstream and downstream of the phantom. The fibers are grouped into 32 bundles of 12 fibers each and one fiber from each bundle is fed into one of 12 silicon photomultiplier tubes (SiPM). The energy detector measures the residual energy of each proton with a 13 cm scintillating block and 16 photomultiplier tubes (PMT) on the backend of the detector. An analog electronics board sums the signals from the 16 PMTs into four different weighted signals which are digitized by the data acquisition system and used for the residual range determination. The compact size of the energy detector requires the protons to have a residual range of less than 10 cm. This means that images with large variations in WET to be imaged with multiple incident proton energies. For example, the pediatric head phantom presented in Section 3.3 has variations in WET between 0 cm and 20 cm and requires three scans of different energies to ensure that each (X, Y) position in the image has protons that pass completely through the phantom and also stop in the scintillating block. For more detailed information on the detector hardware and calibration procedure, we refer the reader to the technical note by DeJongh, et al[9].

### 2.2 Low Intensity Pencil Beam Delivery for Imaging

The proton radiography detector relies on low-intensity pencil beams to track individual protons through the object. An accelerator plan sets the beam delivery conditions and specifies the (X, Y) position of each pencil beam at the isocenter plane, a dwell time for each spot position, and the incident proton energy for the transverse scan. For an image requiring multiple energies, we create a unique accelerator plan for each energy scan. The pencil beams used for these radiographic images acquired at NMCPC have an energy dependent transverse size with sigma in air at isocenter ranging from 0.4 cm to 0.8 cm. The accelerator plan places pencil beam spots spaced 0.5 cm apart on the isocenter plane, dwelling for 10 ms in each spot with an integrated proton flux between 1 and 2 million protons per second. Therefore, a single energy scan with a field size of 20 x 20 $cm^2$ at the isocenter plane delivers between 16 and 32 million protons, which results in a dose of 0.06 to 0.12 mGy.[9] To achieve the low beam intensity, we reduced the proton current at the cyclotron source as well as increased the beamline collimation of the momentum and divergence slits. The cyclotron current and the collimator slit positions are then maintained constant for each energy during the imaging runs.

### 2.3 Reconstruction Methods

Ordoñez, et al[10] previously described the reconstruction methods used in this paper. For clarity, we discuss the major components of the methods here. The proton path is influenced by multiple Coulomb scattering and is accounted for in the reconstruction with a most likely path (MLP) algorithm[11]. The algorithm takes the position of the proton at the upstream and downstream tracker planes as well as the proton's incident angle on the upstream tracker plane to calculate a curved proton trajectory, known as the most likely path, through the object. We reconstruct the image in a plane located at isocenter and segmented into 1 $mm^2$ pixels. The MLP determines the intersection of the proton trajectory with the image plane and a histogram for each pixel accumulates the WEPL values for the protons binned into that pixel. We assign the WET for the pixel as the mean WEPL value. In regions with rapid variations, the WET value may be an average of two or more WEPL peaks.

### 2.4 WET Accuracy

We measured the WET accuracy of the system with a proton radiograph of the custom peg phantom, shown in Figure 2a. The phantom consists of eight 4 cm thick inserts of various tissue-equivalent materials with known RSP and WET values. Each insert has a diameter of 1.8 cm with the WET of the inserts ranging from 0.800 cm to 7.020 cm. The phantom has a background made from blue wax bolus and is 4 cm thick. We independently measured the WET values of each insert by looking at the pull-back of a 150 MeV beam with and without the material in the way using a multi-layer ionization chamber (IBA, Louvain-La-Nueve, Belgium). These values are considered our "true" values and are listed in Table 1.

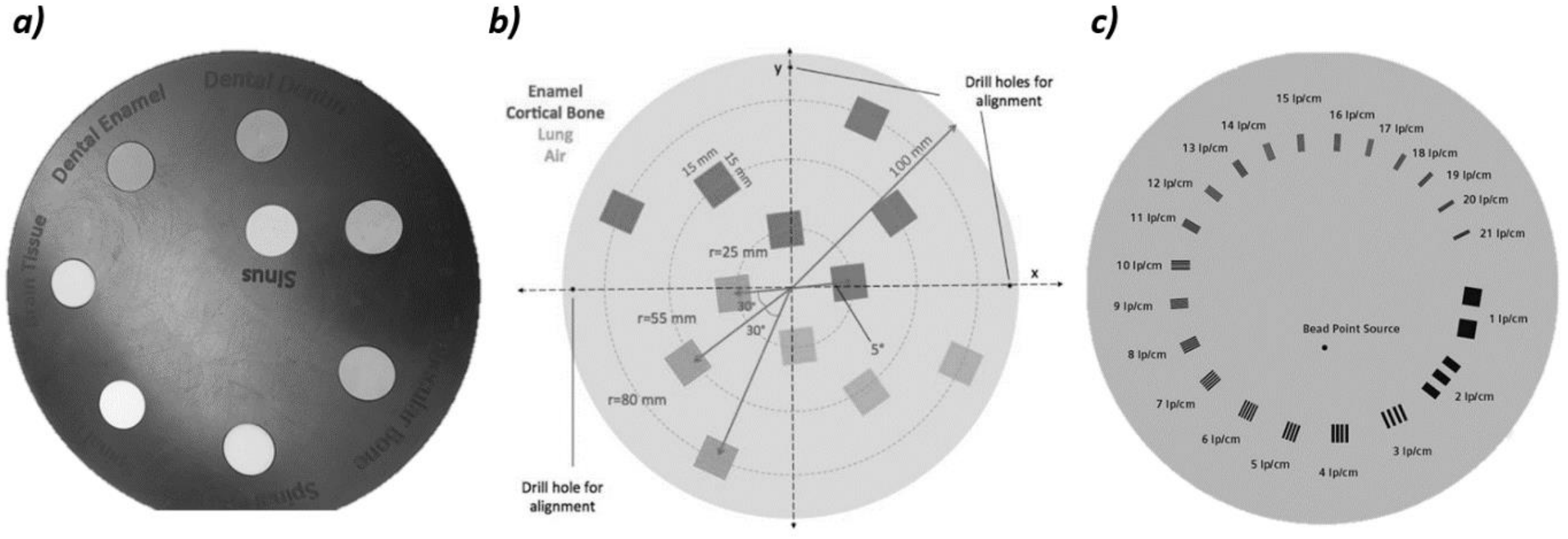

*Figure 2: a) Photo of custom phantom used for WET accuracy measurements. The phantom is 4 cm thick with eight inserts of tissue equivalent materials with known WET values ranging from 0.800 cm to 7.02*

*cm. b) Schematic of the edge phantom designed by Plautz, et al for measuring spatial resolution[12]. The inserts are enamel, cortical bone, lung, and air in a water-equivalent background material. Reprinted with permission from John Wiley and Sons, Inc. c) Schematic of the CATPHAN 528 line pair phantom. The line pairs are aluminum in an epoxy background.*

### 2.5 Spatial Resolution

In proton radiography, spatial resolution is dependent on several parameters due to multiple Coulomb scattering. Therefore, we chose to measure the spatial resolution of the proton radiography system by calculating the modulation transfer function (MTF) on two separate phantoms in different setups. The first, shown in Figure 2b, is a custom edge phantom designed by Plautz, et al[12] with 12 inserts of varying composition and WET. The thickness of the phantom is 6 cm with 4 cm thick inserts. We calculated the MTF by taking the Fourier transform of the derivative of the edge spread function. To accomplish this, we take several line profiles across the edge of the insert. We then overlay these line profiles and align them to the 50% point of the falloff between the background and insert. A sigmoid curve is fit to the points to create an edge spread function. We then take the derivative of the edge spread function to create a Gaussian shaped line spread function. We then perform a one-dimensional Fourier transform on the line spread function. The normalized coefficients of the Fourier transform plotted against the spatial frequency gives the MTF.

A second method for evaluating the MTF used a CATPHAN 528 Line Pair phantom (The Phantom Laboratory Incorporated, Salem, New York), shown in Figure 2c. This phantom consists of aluminum line pair inserts of varying widths and separations set in an epoxy background. During imaging, we place the 4 cm thick phantom in front of approximately 20 cm of solid water for a total background WET of about 24 cm. The aluminum inserts create an additional WET of about 2.5 mm. We calculate the MTF for each line pair (LP) from the maximum and minimum WET values in a profile through each line pair, similar to the method described in Giacometti, et al.[13]:

$$MTF_{LP} = \frac{\langle max(WET)\rangle - \langle min(WET)\rangle}{[\langle max(WET)\rangle - \langle min(WET)\rangle]_{LP=1\ lp/cm}}. \quad (5)$$

In this definition, the MTF for 1 lp/cm is unity. The MTF-10%, which is the spatial frequency at which the MTF value falls to 0.1, defines the spatial resolution of the setup[12].

### 2.6 Relative WET Changes for Patient QA

One clinical application of proton radiography is the ability to detect inter-fractional anatomical changes in the patient that could cause errors in the intended proton range. We imitated a patient with changing internal anatomy with a customized CIRS HN-715 pediatric head phantom (Computerized Imaging Reference Systems, Norfolk, VA) and tested the proton radiograph's capability to be used as a tool to detect changes in WET. This phantom has a 4 x 4 x 4 $cm^3$ cubic cavity with space for a 4 x 4 x 2 $cm^3$ insert sandwiched between two 4 x 4 x 1 $cm^3$ blue bolus wax spacers as shown in Figure 3. We interchanged seven different inserts of various tissue-equivalent plastic materials with relative stopping powers (RSP) that were independently measured with a multi-layer ionization chamber to create known changes in WET. The inserts can be rotated 90 degrees such that radiographs can be taken in the anterior-posterior (AP) direction or the lateral direction. For the first radiograph taken, we filled the phantom with an insert of brain-equivalent material. For subsequent radiographs, we interchanged the insert with another tissue-equivalent material while the base of the phantom was held in place. We created difference maps between radiographs and measured the change in WET by taking the mean and standard error in an ROI of the insert.

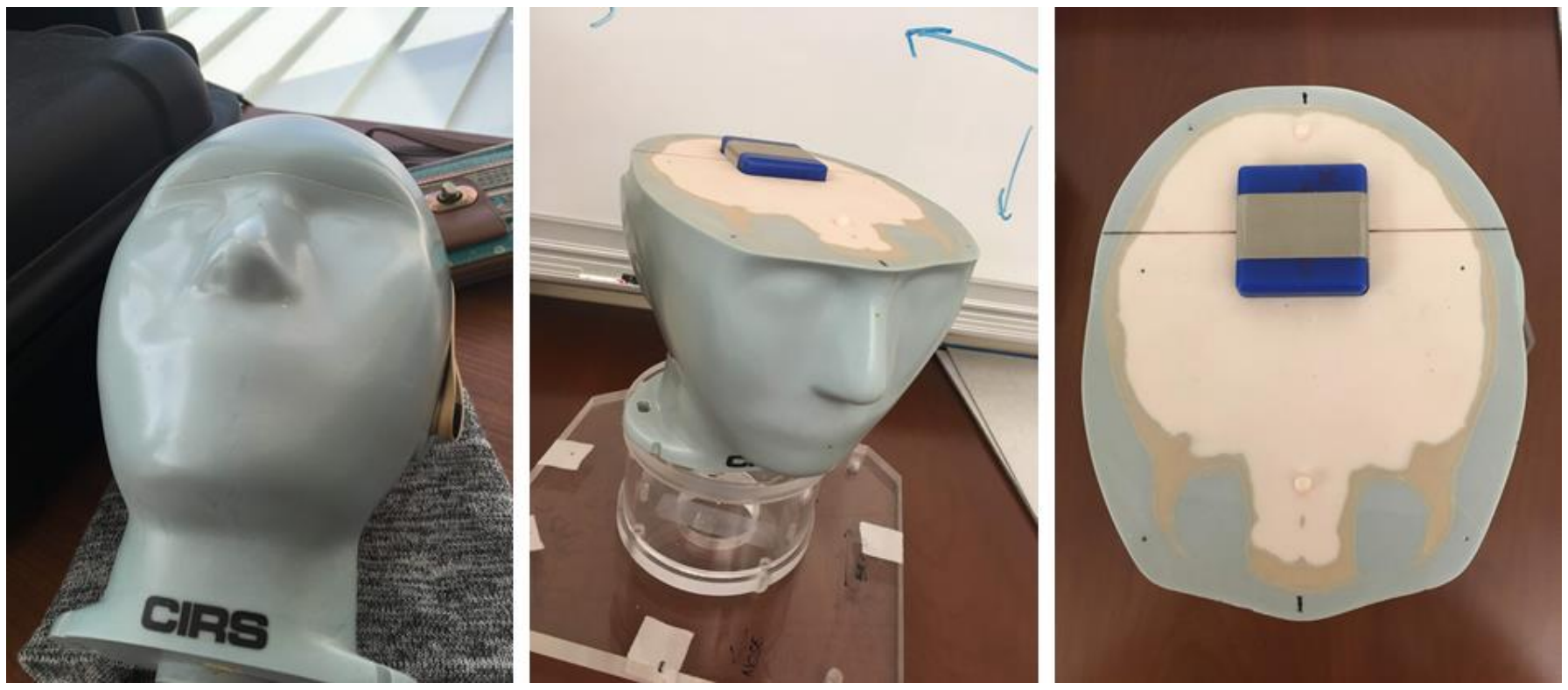

*Figure 3: Photo of the customized pediatric head phantom with tissue-equivalent insert and blue bolus wax spacers used for testing the detection of anatomical changes. In these photos, the insert and spacers are oriented for an AP radiograph. The insert and spacers would be rotated 90 degrees for a lateral radiograph.*

## 3. RESULTS

### 3.1 WET Accuracy

Figure 4 shows a proton radiograph of the custom peg phantom taken using three proton energies (140 MeV, 120 MeV, and 100 MeV). We measured the WET by taking the mean and standard error of the pixel values in a 100-pixel ROI located in each of the inserts. Table 1 shows the results.

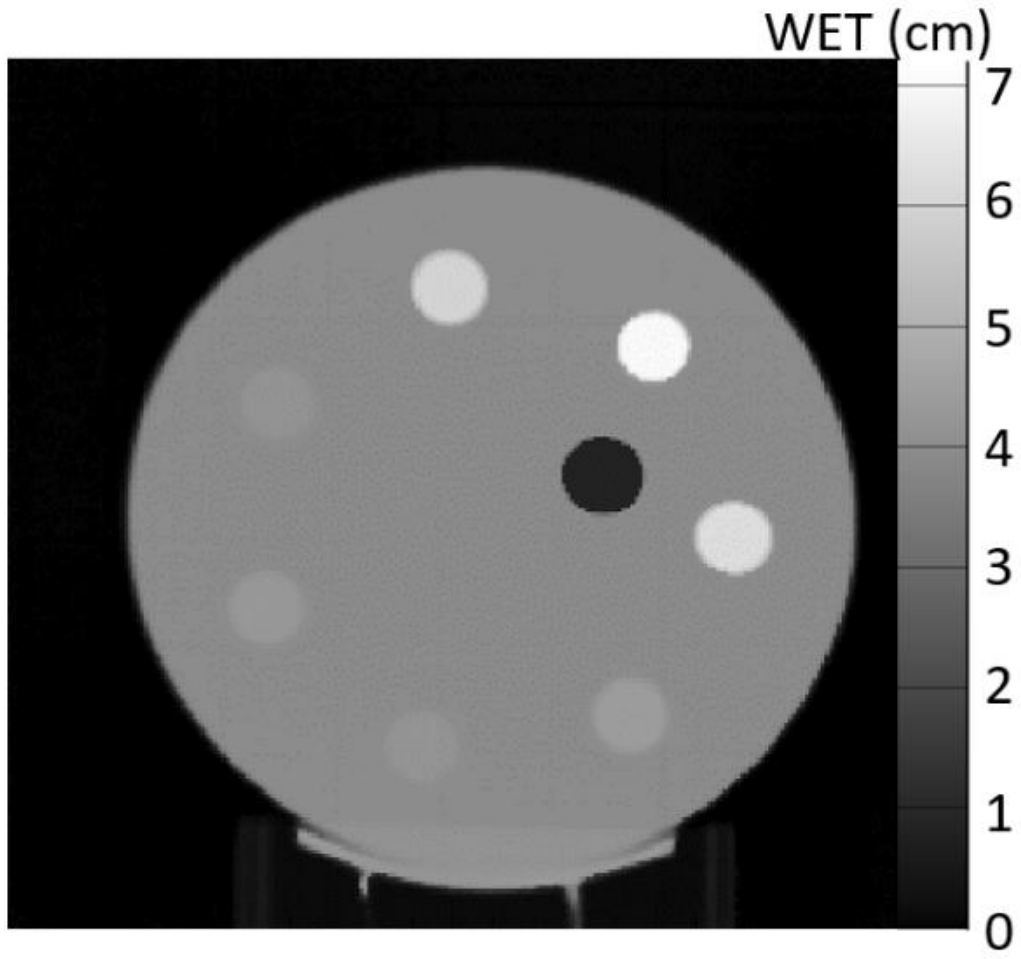

*Figure 4: Proton radiograph of the custom peg phantom shown in Figure 2a. The radiograph was taken with three energies.*

| Insert | True WET (cm) | Measured WET, mean ± standard error (cm) | Difference between True and Measured (cm) | Percent error (%) |
|---|---|---|---|---|
| Sinus | 0.80 | 0.80 ± 0.01 | 0.00 | 0.00 |
| Enamel | 7.02 | 7.01 ± 0.01 | -0.01 | -0.14 |
| Dentin | 5.98 | 5.96 ± 0.01 | -0.02 | -0.33 |
| Brain | 4.16 | 4.15 ± 0.01 | -0.01 | -0.24 |
| Spinal Cord | 4.16 | 4.17 ± 0.01 | 0.01 | 0.24 |
| Spinal Disc | 4.28 | 4.28 ± 0.01 | 0.00 | 0.00 |
| Trabecular Bone | 4.40 | 4.41 ± 0.01 | 0.01 | 0.22 |
| Cortical Bone | 6.22 | 6.21 ± 0.01 | -0.01 | -0.16 |

*Table 1: The measured WET of each of the inserts in the custom phantom*.

**3.2 Spatial Resolution**

Figure 5 shows a proton radiograph of the edge phantom and the MTF corresponding to the four most central inserts. The MTF-10% of each of the inserts range from 0.76 lp/mm to 1.13 lp/mm. The spatial resolution of the four inserts are as follows: 0.76 lp/mm (Lung), 1.13 lp/mm (Air), 0.92 lp/mm (Enamel), and 0.77 lp/mm (Cortical).

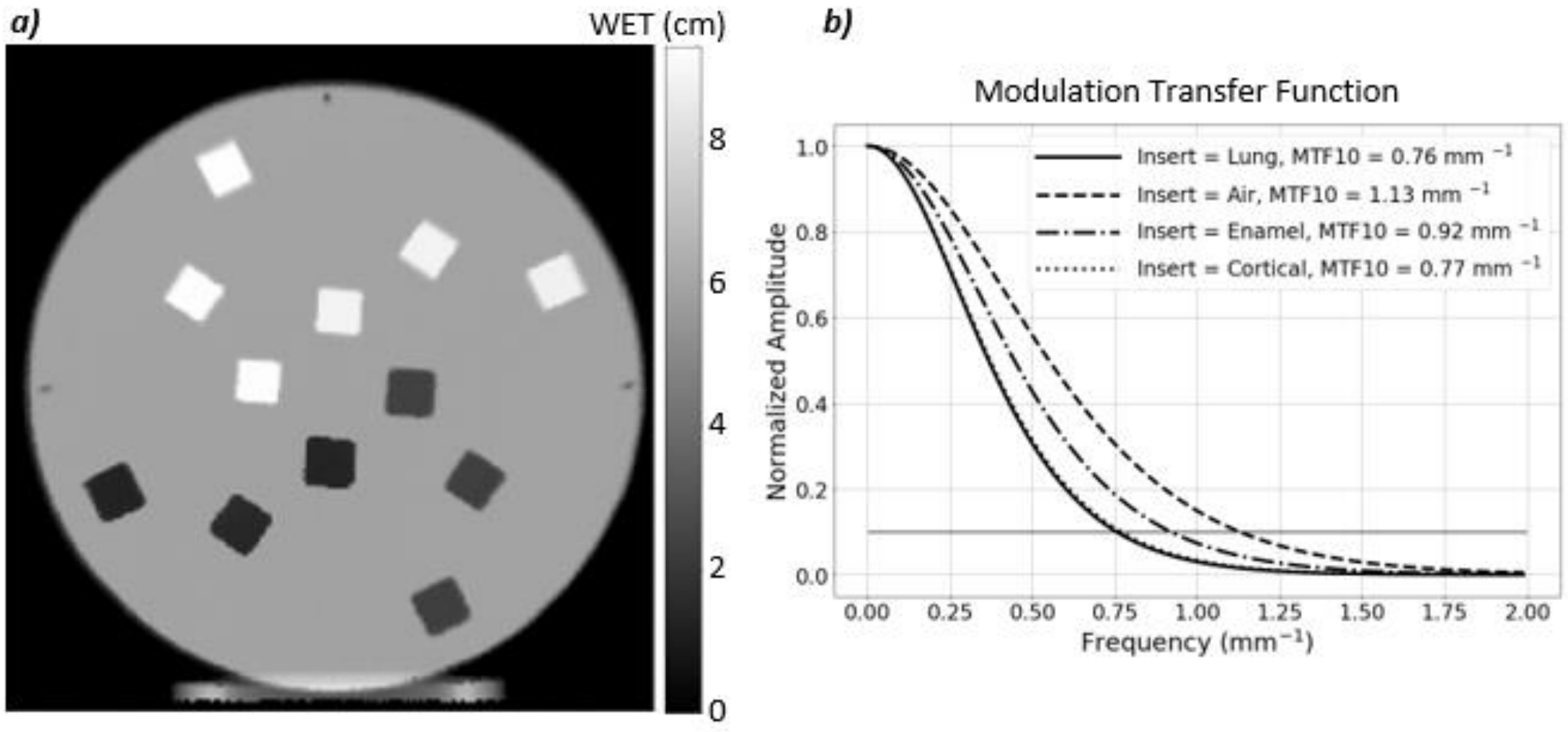

*Figure 5: a) Proton radiograph of the edge phantom. b) The MTF plot of the four most central inserts. The MTF-10% for each of the inserts are shown in the top right corner of the MTF plot.*

Figure 6 shows a proton radiograph of the line pair phantom and the corresponding MTF. The MTF-10% is equal to 0.6 line pairs per millimeter as seen in Figure 6b.

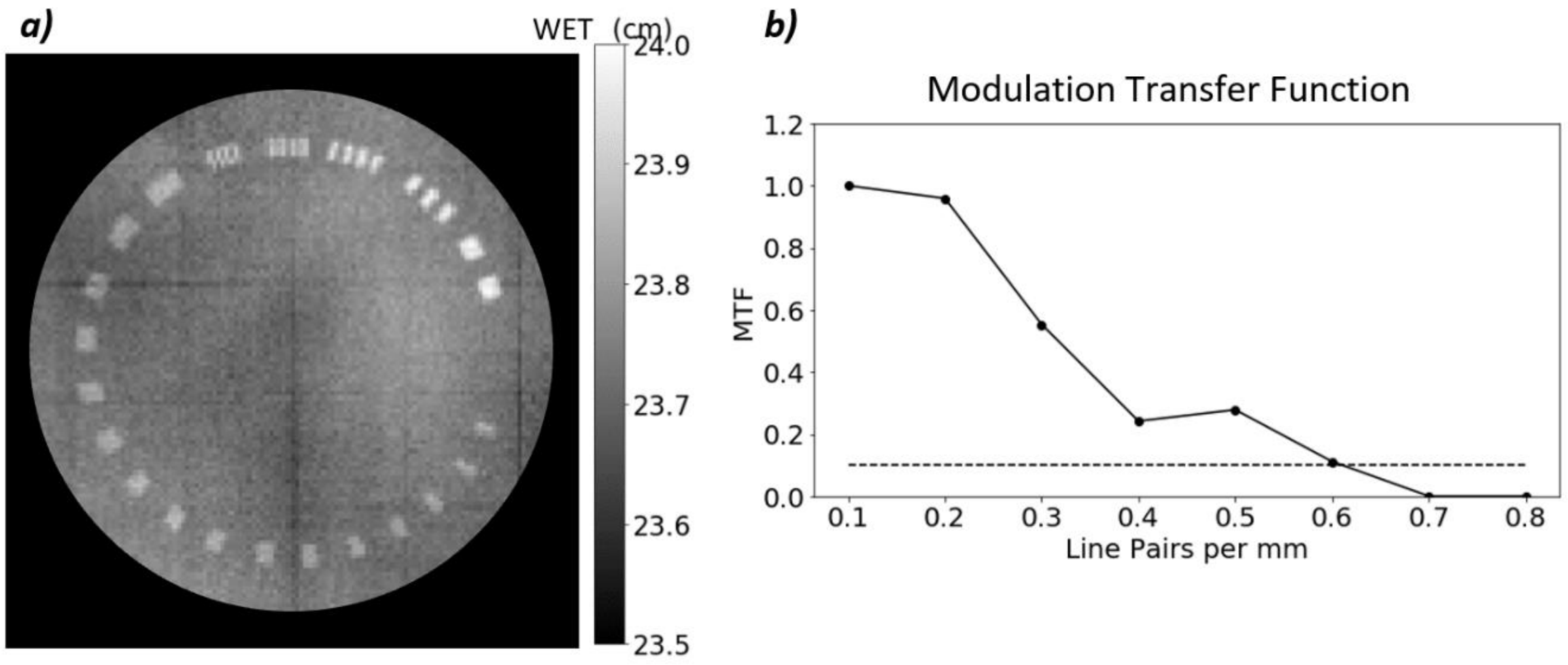

*Figure 6: a) Proton radiograph of CATPHAN 528 line pair phantom. The grid artifacts visible in the image result from nonuniformities in the tracker plane construction and will be eliminated in subsequent constructions. They are visible here due to the reduced WET scale. b) MTF plot of the line pairs. The MTF-10% of the line pair phantom is 0.6 lp/mm.*

### 3.3 Relative WET Differences for QA

Figure 7 shows a series of proton radiographs with various tissue-equivalent inserts in the customized CIRS pediatric head phantom. We created difference maps, such as those in Figure 8a, between the radiograph with the brain insert and the radiographs with the other inserts. We measured the difference in WET in the region of the insert from a 600-pixel ROI in the center of the insert and avoiding the gap between the superior and inferior parts of the phantom. Table 2 shows the results of the WET difference measurements for all inserts. The first column indicates the insert we are comparing to brain and the second column lists the true WET difference between the inserts. The third and fifth columns report the mean and standard error of the measured WET difference for AP and lateral radiographs, respectively. The fourth and sixths columns report the error between the true WET difference and the measured WET difference and the percent error between measurement and truth as a function of the total range.

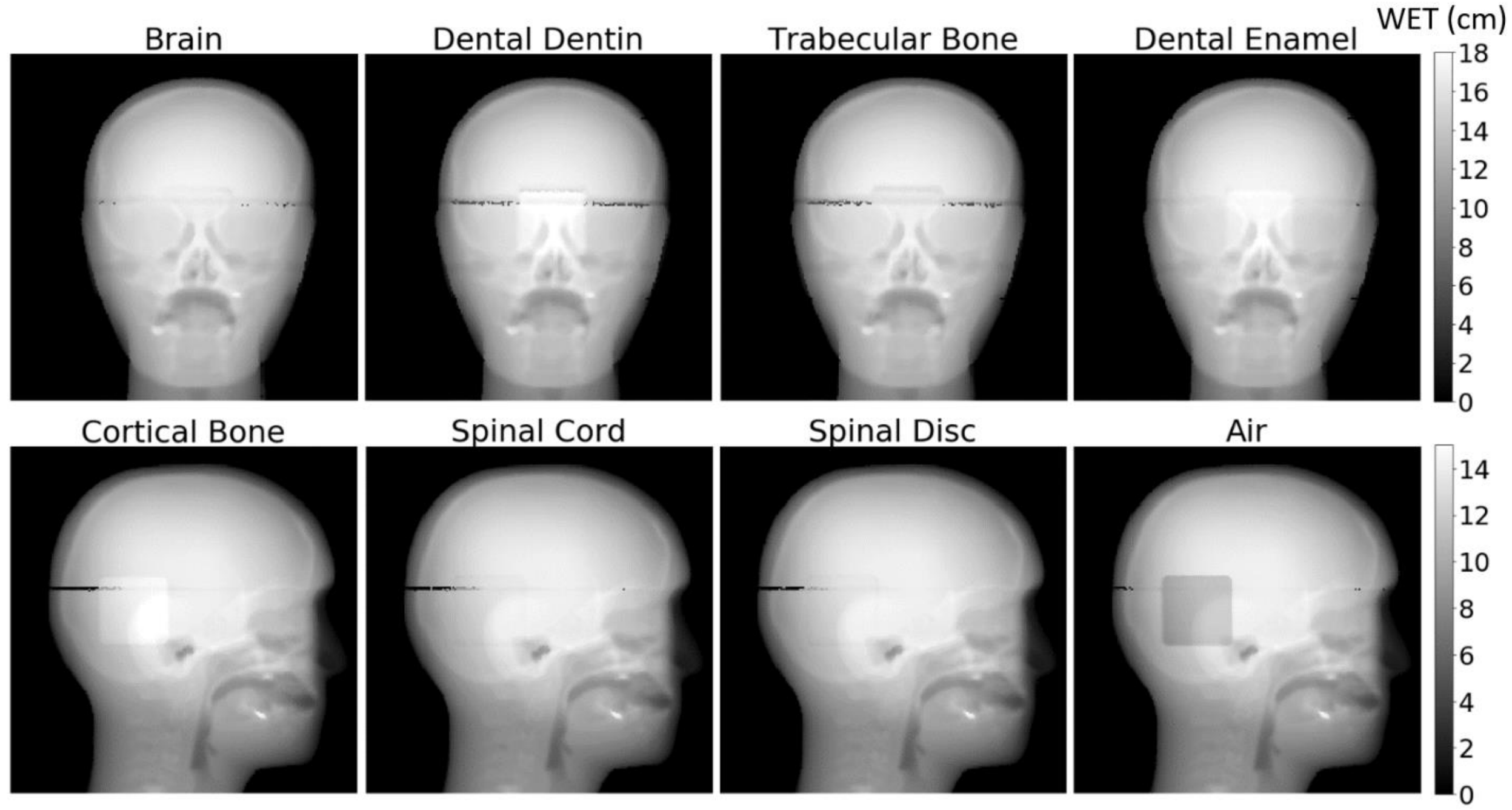

*Figure 7: AP and lateral proton radiographs of the customized pediatric head phantom with the various inserts. A higher density insert, such as cortical bone, will have a higher WET and therefore will appear brighter on the radiograph. A lower density insert, such as air, will have a lower WET and therefore appear darker on the radiograph*. The dark artifact running across the phantom is the air gap where the superior and inferior portions of the phantom come together.

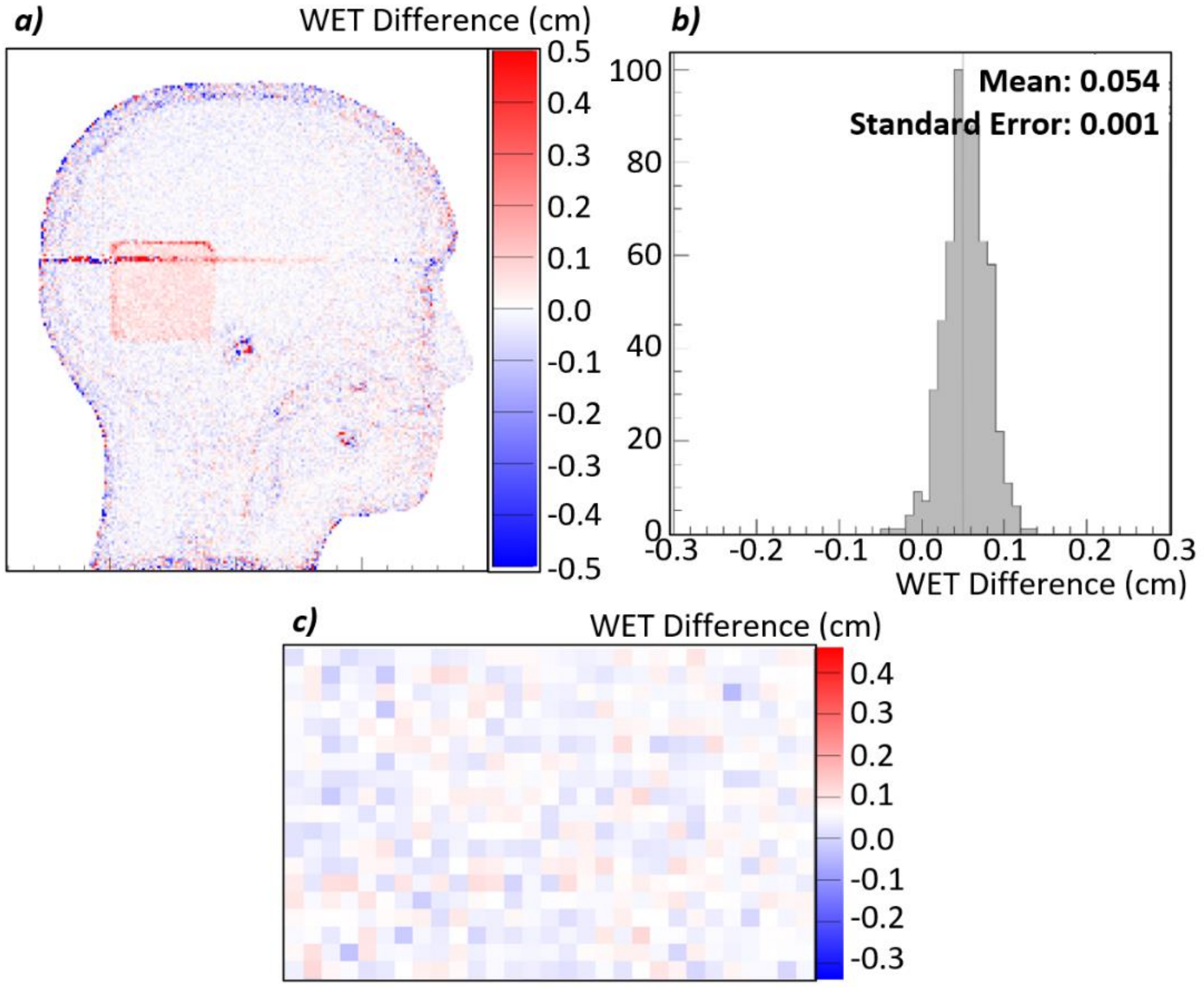

*Figure 8: a) Difference map between a proton radiograph of the phantom with the brain insert and a proton radiograph of the phantom with the spinal disc insert. The red color indicates an increase in WET from brain to spinal disc, while blue indicates a decrease. In this case, the spinal disc insert has a higher WET than the brain insert. b) Histogram of the WET difference within the 600-pixel ROI. The mean of the distribution is 0.054 cm with a standard error of 0.001 cm. c) Enlarged view of the ROI. The color is scaled such that pixels where the WET differences, true minus measured, appears white. The even distribution of red and blue indicates that the noise is consistently distributed throughout the large ROI.*

| Insert | True WET Difference (cm) | Measured AP WET Difference, mean ± standard error (cm) | AP Measurement Error (cm) / Percent Difference of total range (%) | Measured Lateral WET Difference, mean ± standard error (cm) | Lateral Measurement Error (cm) / Percent Difference of total range (%) |
|---|---|---|---|---|---|
| Air | -2.080 | -2.073 ± 0.003 | 0.007 / 0.05 | -2.073 ± 0.002 | 0.007 / 0.07 |
| Spinal Cord | 0.000 | 0.009 ± 0.002 | 0.009 / 0.06 | 0.005 ± 0.001 | 0.005 / 0.04 |
| Spinal Disc | 0.060 | 0.065 ± 0.002 | 0.005 / 0.03 | 0.054 ± 0.001 | -0.006 / -0.05 |
| Trabecular Bone | 0.120 | 0.130 ± 0.003 | 0.010 / 0.06 | 0.118 ± 0.002 | -0.002 / -0.02 |
| Dentin | 0.910 | 0.936 ± 0.003 | 0.026 / 0.15 | 0.900 ± 0.001 | -0.010 / -0.08 |
| Cortical Bone | 1.030 | 1.052 ± 0.002 | 0.022 / 0.13 | 1.020 ± 0.001 | -0.010 / -0.08 |
| Dental Enamel | 1.430 | 1.453 ± 0.003 | 0.023 / 0.13 | 1.423 ± 0.002 | -0.013 / -0.05 |

*Table 2: Table describing the measured WET differences between two radiographs of a pediatric head phantom with various tissue-equivalent inserts. The second column shows the true WET difference between a brain-equivalent insert and another tissue-equivalent insert. The third and fifth column show the measured difference from difference maps created between the two proton radiographs and their respective standard errors. The fourth and sixth columns show the physical difference between the truth and measured WET difference and the percent difference as a function of the total range (12.04 cm and 16.32 cm, respectively) calculated by* $\frac{(TotalRange + Measured) - (TotalRange + True)}{TotalRange + True}$.

## 4. DISCUSSION

### 4.1 Spatial Resolution

Accurate discernment of the edges of the anatomical structures requires appropriate spatial resolution. The line pair phantom study showed a spatial resolution of 0.6 lp/mm and the edge phantom study showed spatial resolution as high as 1.13 lp/mm. Unlike with x-ray radiography, spatial resolution has a complex relation with many parameters due to the multiple Coulomb scattering. Spatial resolution in proton radiography is a function of the position of the object inside the phantom, the detector positions, and the thickness of the phantom, among others. We report spatial resolution under very specific conditions as defined in the text. The lower spatial resolution measured by the line pair phantom setup may be attributed to the larger amount of material the proton must pass through as compared to the edge phantom (24 cm and 6 cm, respectively). Additionally, the location of the image plane relative to the inserts may affect the spatial resolution as well. A systematic study is necessary to determine the setup factors that most affect the spatial resolution.

### 4.2 WET Accuracy and Detection of Anatomical Changes

WET accuracy of a proton radiograph allows for accurate proton range predictions immediately prior to treatment. Because proton radiography is a direct measurement of the WET through the patient, any changes in anatomy that could cause a change in the proton range is directly measured. Common clinical guidelines include an additional ~3.5% error margin on the distal and proximal edge of the tumor due to uncertainties in the proton range introduced during treatment planning[14]. We measured the accuracy of the proton radiographs to better than ±0.02 cm or 0.33% and the difference measurements to better than ±0.023 cm or 0.15% of the measured values. These values are on the order of accuracies measured by Krah et al.[15] with an alternative radiography system. Additionally, the accuracy is well within the 3.5% planning margin, so the clinical benefits for reducing distal margins can be considered.

We detected WET changes as small as 0.6 mm. The WET difference measured on the lateral images have better accuracy than those measured on the AP images. This is due to more heterogeneities in the AP direction than in the lateral direction. In both cases, the detected WET changes are smaller than the typical 3.5% distal range margin included during treatment planning in proton therapy. This test implies that proton radiography could be used as a tool to detect anatomical changes that would cause errors in proton range and the dose distribution. To enhance this study further, smaller objects that are more spherical in shape could be used to better represent human anatomy. Furthermore, live tissue samples could be better for this test. However, the WET of live tissues is much more difficult to accurately quantify. We expect the results from the tissue-equivalent materials will be very similar to real tissues.

## 5. CONCLUSION

The pre-clinical proton radiography system developed by ProtonVDA LLC showed good image quality and usefulness for detection of anatomical changes in a patient prior to each daily treatment. The spatial resolution measured from the MTF-10% was between 0.6 lp/mm and 1.13 lp/mm, and the WET accuracies for the various inserts were ±0.2 mm or 0.33%.

Clinically, proton radiography has the capability to detect proton range differences due to anatomical changes in a patient. Difference maps between proton radiographs of a pediatric head phantom with varying inserts show that proton range differences as low as 0.6 mm can be detected to within 0.13% of the true value, well within the 3.5% planning margin.

## ACKNOWLEDGEMENTS

This work used resources of the Center for Research Computing and Data at Northern Illinois University and resources at Northwestern Medicine Chicago Proton Center. The authors thank Reinhard Schulte, MD from Loma Linda University for reviewing and commenting on this paper. We also thank Nick Detrich from Ion Beam Applications for his work to create control software to deliver proton spot patterns at the correct intensity and energies required by the radiography system. In addition, we thank Igor Polnyi for his help assembling the detector and assisting during data collection.

This work was sponsored by the National Cancer Institute of the National Institutes of Health contract numbers R44CA203499 and R44CA243939, the US Department of the Army contract number W81XWH-10-1-0170, and the US Department of Energy contract number DE-SC0005135. The US Army Medical Research Acquisition Activity, 820 Chandler Street, Fort Detrick MD 21702-5014 is the awarding and administering acquisition office for contract number W81XWH-10-1-0170. The content in this article does not necessarily reflect the position or policy of the Government, and no official endorsement should be inferred.

**CONFLICT OF INTEREST STATEMENT**

The authors have intellectual property rights to the innovations described in this paper. James S. Welsh has served as a medical advisor to ProTom International. Don F. DeJongh and Victor Rykalin are co-owners of ProtonVDA, LLC.

**FOOTNOTES**

a. WEPL is slightly different from WET. A block of material has a fixed WET value defined by the physical thickness (in beam direction) and the relative stopping power of the materials. WEPL is defined for each proton as the path integral of the RSP along the individual trajectory of the proton. Averaging over many proton WEPL values gives an estimate of WET.